\documentclass[12pt]{iopart}
\usepackage{iopams}

\begin{document}

\paper{An exactly solvable model for a $\beta$-hairpin with random interactions}

\author{Marco Zamparo}
\address{Dipartimento di Fisica, INFN sezione di Torino and CNISM,\\ 
Politecnico di Torino, Corso Duca degli Abruzzi 24, Torino, Italy}
\ead{marco.zamparo@polito.it}

\begin{abstract}
I investigate a disordered version of a simplified model of protein folding, with binary degrees
of freedom, applied to an ideal $\beta$-hairpin structure. Disorder is introduced by assuming that
the contact energies are independent and identically distributed random variables.
The equilibrium free-energy of the model is studied, performing the exact calculation of its quenched 
value and proving the self-averaging feature.
\end{abstract}

\maketitle

\section{Introduction}

The present paper is devoted to the analysis of a simple disordered model for an ideal $\beta$-hairpin
structure, for which some exact results may be derived. Disordered models originate very intricated
scenario and their study needs new mathematical methods and algorithms; reffering to plain models with a 
known solution could be helpful to test them.  

The model I consider is a disordered version of one introduced by Wako and Sait\^{o} \cite{WS1,WS2} 
in 1978 and independently reintroduced by Mu\~{n}oz and co-workers \cite{ME1,ME2,ME3} in the late 90's 
to inquire into the problem of protein folding. The Wako-Sait\^{o}-Mu\~{n}oz-Eaton (WSME)
model is a highly simplified one where the purpose 
is describing the equilibrium of the protein folding process under the assumption that it is mainly determined 
by the structure of the native state (the functional state of a protein), whose knowledge is assumed. It is a 
one-dimensional model, with long-range, many-body interactions, where a binary variable is associated to each
peptide bond (the bond connecting consecutive aminoacids), denoting the native and unfolded conformation.
Two aminoacids can interact only if they are in contact in the native state and all the peptide bonds between 
them are ordered. Moreover an entropic cost is associated with each ordered bond. 

Many papers have been published in the last few years concerning the equilibrium properties of the model
and its exact solution \cite{Amos,BP,P}, its kinetics \cite{ZP1,ZP2,BPZ1} and some generalizations 
to the problem of mechanical unfolding \cite{IPZ1,IPZ2}.
In particular in \cite{Amos} the exact solution for a homogeneous $\beta$-hairpin structure was given,
while in \cite{BP} one can find the exact treatment in the general case.
Recently the model has been applied to the analysis of real proteins 
\cite{IS1,IS2,HE,IS3,AW,BPZ2,IP,ZP3} and, rather interestingly, in a problem of strained epitaxy 
\cite{TD1,TD2,TD3}.

In order to introduce some disorder in the WSME model, I suppose the contact energies are independent
quenched variables. This assumption has been done for the base pairing energies in some models for the
ribonucleic acid (RNA) secondary structure \cite{Hwa}, where one aims at retaining the spirit of Watson-Crick pairing
that interactions between some specific bases are favoured with respect to the others.
However, even if the $\beta$-hairpin structure mimics the zipper features of the RNA secondary structure, 
the purpose of this paper is the modest one of proposing a simple exactly solvable disordered model,
calculating the free-energy and proving its self-averaging property. 
The computation of the quenched free-energy,  i.e. the average of the 
free-energy over the quenched disorder, will be provided avoiding the replica theory \cite{EA} and making 
use of some properties of the free-energy itself which will be proven rigorously in advance.

The paper is organised as follow: in Section \ref{model} the WSME model and its disordered version for the 
$\beta$-hairpin structure are introduced. Section \ref{free} is devoted to the calculation of the quenched 
free-energy and Section \ref{self} to prove self-averaging. Conclusions are drawn in Section 
\ref{conclusions}.

\section{The model}
\label{model}

The WSME model describes a protein of $N+1$ residues as a chain of $N$ peptide bonds connecting
consecutive aminoacids. In order to identify the native (ordered) conformation and distinguish it
from the unfolded (disordered) one, a binary variable $m_k$ is associated to the peptide bond $k$, $k=1,\ldots,N$. 
Each variable, related to the values of the dihedral angles at the same peptide bond, assumes value 1 in
the native state and 0 otherwise. 
Since the unfolded state allows a much larger number of microscopic realizations than the native one, an 
entropic cost $q_k$ is given to the ordering of the peptide bond $k$.
The main assumption about the interactions is that two bonds can interact only if they are in contact in the 
native state (so that the model can be classified as G\={o}-like \cite{Go}) and all bonds between them are ordered.  

The Hamiltonian of the model (an effective free-energy, properly speaking) reads
\begin{equation}
H_N(m)=\sum_{i=1}^{N-1}\sum_{j=i+1}^{N}\epsilon_{ij}\Delta_{ij}\prod_{k=i}^{j}m_k + k_BT\sum_{k=1}^{N}q_km_k,
\label{Hgeneral}
\end{equation}
where $T$ is the absolute temperature.
The product $\prod_{k=i}^{j}m_k$ takes value 1 if and only if all the peptide bonds going from $i$ to $j$ are 
ordered, thereby realizing the assumed interaction. The contact matrix elements $\Delta_{ij}\in\{0,1\}$ tell
us which bonds are at close distance in the native state. Finally, the contact energies $\epsilon_{ij}<0$ 
quantify the intensity of the contacts.

An ideal $\beta$-hairpin with an odd number $2N+1$ of peptide bonds is characterized by the contact matrix 
elements $\Delta_{ij}$ equal to 1 if $i+j=2N+2$ and 0 otherwise. The structure results in the characteristic 
Hamiltonian (divided by $k_BT$)
\begin{equation}
H_N^{\epsilon}(m)=-\beta\sum_{i=1}^{N}\epsilon_{i}\prod_{k=N+1-i}^{N+1+i}m_k + q\sum_{k=1}^{2N+1}m_k,
\label{H}
\end{equation}
where $\beta=1/k_BT$.

In this work I concentrate on the case in which $\epsilon_1,\ldots,\epsilon_N$ are independent random variables
identically distributed in a set $\cal{E}\subseteq\mathbb{R}$ according to a probability
measure~$P$. Moreover, in order to deal with a homogeneous model having a thermodynamic limit, 
the entropic cost $q_k$ is chosen equal to $q$ for any $k$, as the comparison between the Hamiltonians
(\ref{Hgeneral}) and (\ref{H}) shows.
I shall assume $P$ is any probability measure satisfying the condition 
$\int_{\epsilon\in\cal{E}}\exp{(\beta\epsilon)}P(d\epsilon) < \infty$, given an arbitrary real value of $\beta$,
and from now on I will denote by $\mu$ the expectation of the contact energy and with
$P_N$ the product measure $P\times\ldots\times P$ $N$-times.

Let us denote with $f_N$ the quenched free-energy (times~$\beta$)
\begin{equation}
f_N(\beta,q) = -\frac{1}{2N+1}~\mathbb{E}[\log Z_N] \doteq
-\frac{1}{2N+1}\int_{\epsilon\in{\cal{E}}^N}\log Z_N(\epsilon) ~ P_N(d\epsilon),
\label{fdef}
\end{equation}
where $Z_N(\epsilon)$ is the partition function of the model (\ref{H}) given a sequence 
$\epsilon=(\epsilon_1,\ldots,\epsilon_N)$ of interaction 
energies:
\begin{equation}
Z_N(\epsilon)=\sum_{m\in\{0,1\}^{2N+1}}\exp[-H_N^{\epsilon}(m)].
\label{Zdef}
\end{equation}
$f\doteq\lim_{N\to\infty}f_N$ is the quenched free-energy in the thermodynamic limit.

\section{The free-energy}
\label{free}

In this section I show how to compute exactly the quenched free energy, discussing some of its properties in advance
and then exploiting them to perform the calculation.
Let us start by observing that, due to the features of the model, it is possible to simplify the expression of the 
partition function $Z_N$. Indeed, summing over the binary variables $m_1$ and $m_{2N+1}$
allows to find the iterative equation \cite{Amos}
\begin{equation}
Z_{N}(\epsilon)=(1+\mbox{e}^{-q})^{2}Z_{N-1}(\epsilon)+(\mbox{e}^{\beta\epsilon_N}-1)\mbox{e}^{\sum_{i=1}^{N-1}\epsilon_i-q(2N+1)}
\end{equation}
valid for any $N\in\mathbb{N}$. Joining this relation to the initial condition 
\begin{equation}
Z_{1}(\epsilon)=(1+\mbox{e}^{-q})^{3}+(\mbox{e}^{\beta\epsilon_1}-1)\mbox{e}^{-3q},
\end{equation}
one obtains immediately the expression
\begin{eqnarray}
\nonumber
Z_{N}(\epsilon) & = & (1+\mbox{e}^{-q})^{2N+1}+\\
& + & \sum_{n=1}^N(\mbox{e}^{\beta\epsilon_n}-1)
\mbox{e}^{\beta\sum_{i=1}^{n-1}\epsilon_i-q(2n+1)}(1+\mbox{e}^{-q})^{2(N-n)}.
\label{Z}
\end{eqnarray}
The formula for $Z_N$ can still be slightly reduced, as it is stated by the following proposition.\\
{\bf Proposition 1.} {\it There exist two positive constants with respect to $N$,} $C$ {\it and} $D${\it, such that}
 \begin{equation}
C\left[1+\sum_{n=1}^{N}\frac{\mbox{e}^{\beta\sum_{i=1}^{n}\epsilon_{i}}}{(1+\mbox{e}^{q})^{2n}}\right]
\leq \frac{Z_{N}(\epsilon)}{(1+\mbox{e}^{-q})^{2N+1}}\leq D\left[1+\sum_{n=1}^{N}\frac{\mbox{e}^{\beta\sum_{i=1}^{n}\epsilon_{i}}}
{(1+\mbox{e}^{q})^{2n}}\right].
\end{equation}

Before sketching the proof, in order to deal with more compact formulas in the following, it is convenient to introduce 
the new quantities
\begin{equation} 
\Xi_N^{\beta,\lambda}(\epsilon)=1+\sum_{n=1}^N\mbox{e}^{\beta\sum_{i=1}^{n}\epsilon_{i}-\lambda n}
\label{Xi}
\end{equation}
and
\begin{equation}
g_N(\beta,\lambda)=\frac{1}{N}~\mathbb{E}[\log \Xi_N^{\beta,\lambda}]
\label{g}
\end{equation}
where the explicit dipendence on $\beta$ and $\lambda$ is taken into account, and rewrite $f$ in the form
\begin{equation}
f(\beta,q)=-\log(1+\mbox{e}^{-q})-\frac{1}{2}~g(\beta,2\log(1+\mbox{e}^q))
\label{f}
\end{equation}
with $g\doteq\lim_{N\to\infty}g_N$. The relationship between the free-energy and the model parameters
comes from the evaluation of the function $g$, so that I shall focus on $g$ rather than $f$.\\
{\bf Proof of Proposition 1.} Looking at the expression (\ref{Z}) and splitting the term 
$(\mbox{e}^{\beta\epsilon_n}-1)$ in the sum, it is possible to rewrite $Z_N$ in the following manner:
\begin{eqnarray}
\nonumber
\fl \frac{Z_{N}(\epsilon)}{(1+\mbox{e}^{-q})^{2N+1}} & = & 1-(1+\mbox{e}^q)^{-3}+\\
& + & \frac{1-(1+\mbox{e}^q)^{-2}}{1+\mbox{e}^q}
\sum_{n=1}^{N-1}\frac{\mbox{e}^{\beta\sum_{i=1}^{n}\epsilon_{i}}}{(1+\mbox{e}^{q})^{2n}}+
(1+\mbox{e}^q)^{-1}\frac{\mbox{e}^{\beta\sum_{i=1}^{N}\epsilon_{i}}}{(1+\mbox{e}^{q})^{2N}}.
\end{eqnarray}
The statement of the proposition is achieved by choosing
\begin{equation}
C=\min\biggl\{1-(1+\mbox{e}^q)^{-3},\frac{1-(1+\mbox{e}^q)^{-2}}{1+\mbox{e}^q},(1+\mbox{e}^q)^{-1}\biggr\}>0
\end{equation}
and
\begin{equation}
D=\max\biggl\{1-(1+\mbox{e}^q)^{-3},\frac{1-(1+\mbox{e}^q)^{-2}}{1+\mbox{e}^q},(1+\mbox{e}^q)^{-1}\biggr\}>0.
\end{equation}

Let us now go over the properties of $g$ that shall allow its evaluation. 
From a physical point of view one is interested only in positive
values of $\beta$ and $\lambda$, but for analitycal reasons it is convenient to assume $\beta$ and $\lambda$ taking
any real value. The first property I show concerns the behaviour of $g$ under reflection with 
respect to the origin.\\
{\bf Proposition 2.}  $g(\beta,\lambda)=\beta\mu-\lambda+g(-\beta,-\lambda)$
{\it where} $\mu$ {\it is the expectation value of the energy contact:} 
\begin{equation}
\mu=\int_{\epsilon\in\cal{E}}\epsilon P(d\epsilon).
\end{equation}
{\bf Proof of Proposition 2.} Remembering the definition (\ref{Xi}), we have
\begin{eqnarray}
\nonumber
\fl \Xi_N^{\beta,\lambda}(\epsilon) & = & 1+\sum_{n=1}^{N-1}\mbox{e}^{\beta\sum_{i=1}^n\epsilon_i-\lambda n}+
\mbox{e}^{\beta\sum_{i=1}^N\epsilon_i-\lambda N}\\
\fl & = & \mbox{e}^{\beta\sum_{i=1}^N\epsilon_i-\lambda N}\left[\mbox{e}^{-\beta\sum_{i=1}^N\epsilon_i+\lambda N}+
\sum_{n=1}^{N-1}\mbox{e}^{-\beta\sum_{i=n+1}^N\epsilon_i+\lambda (N-n)}+1\right]
\end{eqnarray}
and changing $n$ with $N-n$ in the sum, we can go on writing
\begin{eqnarray}
\nonumber
\fl \Xi_N^{\beta,\lambda}(\epsilon_1,\ldots,\epsilon_N) & = & \mbox{e}^{\beta\sum_{i=1}^N\epsilon_i-\lambda N}
\left[\mbox{e}^{-\beta\sum_{i=1}^N\epsilon_i+\lambda N}+
\sum_{n=1}^{N-1}\mbox{e}^{-\beta\sum_{i=N-n+1}^N\epsilon_i+\lambda n}+1\right]\\
\nonumber
& = & \mbox{e}^{\beta\sum_{i=1}^N\epsilon_i-\lambda N}
\left[1+\sum_{n=1}^{N}\mbox{e}^{-\beta\sum_{i=N-n+1}^N\epsilon_i+\lambda n}\right]\\
\nonumber
& = & \mbox{e}^{\beta\sum_{i=1}^N\epsilon_i-\lambda N}
\left[1+\sum_{n=1}^{N}\mbox{e}^{-\beta\sum_{i=1}^n\epsilon_{N-i+1}+\lambda n}\right]\\
& = & \mbox{e}^{\beta\sum_{i=1}^N\epsilon_i-\lambda N} ~ \Xi_N^{-\beta,-\lambda}(\epsilon_N,\ldots,\epsilon_1).
\label{int}
\end{eqnarray}
The connection (\ref{g}) between $\Xi_N^{\beta,\lambda}$ and $g_N$ allows us to conclude immediately the proof.

The second result I report describes a homogeneity property of $g$.\\
{\bf Proposition 3.} $g(t\beta,t\lambda)=t g(\beta,\lambda)$ {\it for any} $t>0$.\\
{\bf Proof of Propostion 3.} At first let us suppose $t\geq 1$. From the inequality, valid for $x>0$,
\begin{equation}
(1+x)^t\geq 1 + x^t
\end{equation}
and from the convexity of the function $x\to x^t$, $x>0$, it follows that
\begin{equation}
\sum_{n=0}^{N} a_{n}^{t} \leq \left(\sum_{n=0}^{N} a_{n} \right)^{t} \leq (N+1)^{t-1} \sum_{n=0}^{N} a_{n}^{t}
\end{equation}
for any integer $N$ and positive numbers $a_{0},\ldots,a_{N}$. This chain of inequalities implies
\begin{equation}
\Xi_{N}^{t\beta,t\lambda}(\epsilon) \leq \left[\Xi_{N}^{\beta,\lambda}(\epsilon)\right]^{t} 
\leq (N+1)^{t-1} ~ \Xi_{N}^{t\beta,t\lambda}(\epsilon)
\end{equation}
and then $g(t\beta,t\lambda)=tg(\beta,\lambda)$ when $t\geq 1$. Bearing in mind the latter point, 
the substitution of $\beta$ with $\beta/t$ and $\lambda$ with $\lambda/t$ allows us to prove the 
proposition also when $0<t<1$.

Finally we can easily characterize $g$ in a region of the parameter space.\\
{\bf Proposition 4.} 
$g(\beta,\lambda)=0$ {\it if} $\lambda\geq\log\int_{\epsilon\in\cal{E}}\mbox{e}^{\beta\epsilon}P(d\epsilon)$.\\
{\bf Proof of Proposition 4.} Making use of the concavity of the logarithm function, we obtain
\begin{eqnarray}
\nonumber
0 \leq g_{N}(\beta,\lambda) & \leq & \frac{1}{N} \log\int_{\epsilon\in{\cal{E}}^N}
\Xi_{N}^{\beta,\lambda}(\epsilon)P_N(d\epsilon)\\
& = & \frac{1}{N}\log\left[1+\sum_{n=1}^{N}\left(\mbox{e}^{-\lambda}\int_{\epsilon\in\cal{E}}
\mbox{e}^{\beta\epsilon}P(d\epsilon)\right)^n\right].
\end{eqnarray}
Then $g(\beta,\lambda)=0$
if $\mbox{e}^{-\lambda}\int_{\epsilon\in\cal{E}}\mbox{e}^{\beta\epsilon}P(d\epsilon)\leq 1$
or equivalently $\lambda\geq \log\int_{\epsilon\in\cal{E}}\mbox{e}^{\beta\epsilon}P(d\epsilon)$.

Exploiting these properties, it is now feasible to show the form of the function $g$ for the whole parameter space. 
From proposition 3 and 4 it follows that, given $t$ larger than 0, $g$ vanishes if
$\lambda\geq\frac{1}{t}\log\int_{\epsilon\in\cal{E}}\mbox{e}^{t\beta\epsilon}P(d\epsilon)$.
Taking the limit $t\to 0^{+}$, this condition reduces to $\lambda\geq \beta\mu$.
On the other hand, if $\lambda\leq\beta\mu$ then $-\lambda\geq -\beta\mu$ and proposition 2 tells us
that $g(\beta,\lambda)=\beta\mu-\lambda$, due to the null value of $g(-\beta,-\lambda)$.
Let us conclude by collecting the previous results in a compact formula
by means of the Heaviside function $\theta$ ($\theta(x)=1$ if $x\geq 0$ and 0 otherwise) and 
$\Theta$ defined as $\Theta(x)=x\theta(x)$.
The following holds\\
{\bf Theorem 1.} $g(\beta,\lambda)=(\beta\mu-\lambda)\theta(\beta\mu-\lambda)=\Theta(\beta\mu-\lambda)$.

\section{Self-averaging property}
\label{self}

This section is devoted to the proof of the self-averaging feature of the free-energy.
In order to quantify the fluctuations of the free-energy let us introduce the function $S_N$ defined as
\begin{equation}
S_N(\beta,\lambda)=\mathbb{E}\biggl[\biggl|\frac{1}{N}\log\Xi_N^{\beta,\lambda}-g(\beta,\lambda)\biggr|\biggr].
\end{equation}
As one can easily verify, given a positive number $\delta$, the probability of having a fluctuation larger than or 
equal to $\delta$ is bounded by $S_N$:
\begin{equation}
\mathbb{P}\biggl[\biggl|\frac{1}{N}\log\Xi_N^{\beta,\lambda}-g(\beta,\lambda)\biggr|\geq\delta\biggr]\leq 
\frac{S_N(\beta,\lambda)}{\delta},
\end{equation} 
where the left-hand side is an usual short notation denoting the probability measure of the set
of $\epsilon\in{\cal{E}}^N$ such that 
$g(\beta,\lambda)-\delta\leq \frac{1}{N}\log\Xi_N^{\beta,\lambda}(\epsilon)\leq g(\beta,\lambda)+\delta$.  

The self-averaging property of the free-energy is described by the fact that
$S_N$ vanishes in the thermodynamic limit, as the following theorem states\\
{\bf Theorem 2.} $S(\beta,\lambda)\doteq\lim_{N\to\infty}S_N(\beta,\lambda)=0$ {\it for any real numbers} $\beta$ 
{\it and} $\lambda$.

In order to prove the theorem it is useful to extend to $S$ the reflection result about~$g$.\\
{\bf Proposition 5.} $S(\beta,\lambda)=S(-\beta,-\lambda)$.\\
{\bf Proof of Proposition 5.} From relation (\ref{int}) and proposition 2 we have
\begin{eqnarray}
\nonumber
\fl \frac{1}{N}\log{\Xi_N^{\beta,\lambda}}(\epsilon_1,\ldots,\epsilon_N)-g(\beta,\lambda) & = & 
\beta\biggl(\frac{1}{N}\sum_{i=1}^N\epsilon_i-\mu \biggr) +\\
& + & \frac{1}{N}\log{\Xi_N^{-\beta,-\lambda}}(\epsilon_N,\ldots,\epsilon_1)-g(-\beta,-\lambda),
\end{eqnarray}
which, passing to absolute values and averaging, yields
\begin{equation}
\biggl|S_N(\beta,\lambda)-S_N(-\beta,-\lambda)\biggr|
\leq |\beta| \int_{\epsilon\in{\cal{E}}^N} \biggl|\frac{1}{N}\sum_{i=1}^N\epsilon_i-\mu \biggr|~ P_N(d\epsilon).
\end{equation}
Thanks to the Cauchy-Schwarz inequality we can go on and reach the result
\begin{eqnarray}
\nonumber
\biggl|S_N(\beta,\lambda)-S_N(-\beta,-\lambda)\biggr|
& \leq & |\beta|\sqrt{\int_{\epsilon\in{\cal{E}}^N} \biggl(\frac{1}{N}\sum_{i=1}^N\epsilon_i-\mu \biggr)^2~ P_N(d\epsilon)}\\
& = & \frac{|\beta|}{\sqrt{N}}\sqrt{\int_{\epsilon\in{\cal{E}}}(\epsilon-\mu)^2~ P(d\epsilon)}.
\end{eqnarray}
The proof is concluded considering the limit $N\to\infty$.

Now we can come back to the theorem.\\
{\bf Proof of Theorem 2.} Remembering that $g(\beta,\lambda)=0$ if $\beta\mu\leq\lambda$ and observing that 
$\Xi_N^{\beta,\lambda}(\epsilon)\geq 1$, we have, when $\beta\mu\leq\lambda$,
\begin{equation}
S_N(\beta,\lambda)=\mathbb{E}\biggl[\biggl|\frac{1}{N}\log\Xi_N^{\beta,\lambda}\biggr|\biggr]
=\mathbb{E}\biggl[\frac{1}{N}\log\Xi_N^{\beta,\lambda}\biggr]=g_N(\beta,\lambda)
\end{equation}
and then $\lim_{N\to\infty}S_N(\beta,\lambda)=g(\beta,\lambda)=0$. On the other hand, when $\beta\mu>\lambda$
we obtain from proposition 5 that $S(\beta,\lambda)=S(-\beta,-\lambda)=0$ since $-\beta\mu<-\lambda$.

\section{Conclusions}
\label{conclusions}

In the previous sections we focused on the function $g$, since its study was equivalent to that of the
free-energy $f$. Now we can come back to the expression (\ref{f}) and thanks to the theorem 1 write
the final formula
\begin{equation}
f(\beta,q)=-\log(1+\mbox{e}^{-q}) - \frac{1}{2}~\Theta(\beta\mu-2\log(1+\mbox{e}^q)).
\end{equation}
The free-energy inherits the self-averaging property from $g$ and thus its behaviour is completely
characterized. 

The continuous function $\Theta(x)$ has a discontinuity in the first derivative at
$x=0$ showing that a first order phase transition occurs at the critical value  
$\beta_c(q)=\frac{2}{\mu}\log(1+\mbox{e}^q)$ of $\beta$. This critical point is associated to the 
transition between a disordered phase, the unfolded state of the peptide, and an ordered one, 
the native state, pointing out a two-state behaviour. 

The transition can be better characterized by means of an order parameter $p_N$, function of $\beta$ and $q$, 
measuring the level of the order in the system. We can choose $p_N$ as the thermal and then quenched average 
of the fraction of native bonds.
From definitions (\ref{H}), (\ref{fdef}) and (\ref{Zdef}) it follows the result  
\begin{equation}
p_N(\beta,q)=\frac{\partial f_N(\beta,q)}{\partial q}
\end{equation}
which, passing to the limit $N\to\infty$, allows us to obtain
\begin{equation}
p(\beta,q) \doteq \lim_{N\to\infty}p_N(\beta,q) = \cases{ 1 & \mbox{if $\beta>\beta_c(q)$},\\
\frac{1}{1+\mbox{e}^q} & \mbox{if $\beta<\beta_c(q)$}.} 
\end{equation}
At low temperature, $\beta>\beta_c(q)$, all the peptide bonds are ordered and the protein is in its native state. 
The relationship between $\beta_c$ and the expectation contact energy $\mu$ implies that no ordering
can occur at physical temperature when the interaction is repulsive in average ($\mu<0$).

Let us observe lastly that the free-energy is the same as in a model with no disorder and contact energies fixed 
at the value $\mu$. This means that the quenched disorder does not affect the critical behaviour
and the transition remains sharp of the first order, as in the pure case. This feature could not be considered 
manifest {\it a priori}, since, as far I know, no general result is available for models with 
long-range and many-body interactions in the presence of quenched disorder.

Concluding, in this paper I have studied and solved exactly a simple disordered model, showing at first 
the mathematical expression of the quenched free-energy and then characterising completely the distribution
of the free-energy by proving its self-average feature. The replica trick has been avoided since
a more straightforward way has been found to reach the desired results. I believe these  
might turn out to be helpful as a benchmark for testing methods from disordered system theory, where exact 
solutions are quite rare.

\ack
I am grateful to Alessandro Pelizzola for having stimulated me to produce this work.

\end{document}